\documentclass[%
twocolumn,
 reprint,
groupedaddress,
 amsmath,amssymb,
 aps,
]{revtex4}

\usepackage{graphicx}
\usepackage{dcolumn}
\usepackage{bm}
\usepackage[colorlinks,allcolors=blue]{hyperref}

\begin{document}

\title{Optimal design of error-tolerant reprogrammable  multiport interferometers}

\author{S.A.\,Fldzhyan}
\author{M.Yu.\,Saygin}%
 \email{saygin@physics.msu.ru}


\author{S.P.\,Kulik}
\affiliation{%
 Quantum Technology Centre, Faculty of Physics, Lomonosov Moscow State University, Moscow, Russian Federation
}%












\begin{abstract}
Photonic information processing demands programmable multiport interferometers capable of implementing arbitrary transfer matrices, for which planar meshes of error-sensitive Mach-Zehnder interferometers are usually exploited. We propose an alternative design that uses a single static beam-splitter  and a variable phase shift as the building block. The design possesses  superior resilience to manufacturing errors and losses without extra elements added into the scheme. Namely, the power transmissivities of the static BSs can take arbitrary values in the range from $\approx{}1/2$ to $\approx{}4/5$. We show that the fraction of transfer matrices non-implementable by the interferometers of our design diminishes rapidly with their size.
\end{abstract}



\maketitle

\section{Introduction}

Photonics is progressively playing a more important role in fundamental science and applied areas, motivated by novel developed  approaches to information processing which are well-matched with the qualities of optics. Linear transformations between multiple optical channels are often required by these approaches, thus, making the utilization of multiport interferometer devices a necessity. For example, multiport interferometers are exploited as mode unscramblers~\cite{Unscrambler2017} and parts of photonic neural networks~\cite{Shen2017,Hughes2018}.

In the recent years, optical interferometers have attracted appreciable interest by the quantum information community, because of  the  promising quantum computing platforms that leverage linear-optics  and unique properties of photon discrete variables~\cite{Rudolph2017,Harris2018} and field continuous variables states~\cite{Lobino2018,RalphCV2017}.  Recent works have demonstrated the versatility of linear-optical quantum systems and their ability to perform quantum computing tasks, ranging from the algorithms suggested at the dawn of quantum information theory~\cite{Politi2009,Zhou2013} to more specific ones that disrupt the landscape nowadays, such as the boson sampling algorithm~\cite{Spring2012,Crespi2013,PanBosonSampling} and quantum deep neural networks~\cite{CarolanNeural,Killoran2018}. 

Universal interferometers can be reprogrammed to implement an arbitrary linear transformation defined by a specific transfer matrix. To construct these interferometers, decomposition methods are used that represent unitary matrices as products of simpler building blocks \cite{Reck1994,Clements2016,RobustArchitecture2019}. Among these methods, the most practical are planar decompositions, since they well suit fabrication by the mature techniques of integrated photonics enabling massive production of sophisticated optical circuits~\cite{Harris2018,Shen2017,Taballione:18}. Today, the most often used methods are those proposed by Reck et al in \cite{Reck1994} and Clements et al in \cite{Clements2016}, which decompose unitary matrices into  planar meshes of variable beam-splitters (BSs), having triangular and rectangular forms, respectively. In these schemes, each variable BS is conveniently realized by a standard element of the Mach-Zehnder interferometer (MZI), made up of  two static balanced BSs with variability provided by two phase shifts. Thus, the overall scheme is reprogrammed by  setting the phase shifts~\cite{Taballione:18,Dyakonov2018}.

For these schemes to be universal, it is crucial that the static BSs should  be balanced. However, this condition can not be fully satisfied because of the errors that occur at realization, limiting the scheme's universality~\cite{Mower:15,Burgwal:17}. The negative effect of the errors progresses  as the interferometer size scales up, effectively imposing stringent requirements on the fabrication tolerances and making challenging the realization of large  interferometers.

Methods exist that can restore the universality of the MZI-based schemes at the cost of adding extra elements into their optical schemes~\cite{Miller:15,Burgwal:17}. However, the common drawback of these methods is the burden of auxiliary control needed to manipulate the additional MZIs and the increased real estate occupied by the scheme on the chip. Therefore, developing more efficient designs of error-tolerant inteferometers is highly demanded nowadays. Here, we propose a new design of planar interferometers, which is error-tolerant to manufacturing errors and universal except for a fraction of matrices which rapidly diminishes with their size.

\section{The MZI-based and BS-based  interferometers}\label{sec:MZI_BS}

An $N$-port interferometer may be described by an $N\times{}N$ transfer matrix acting on vectors of field amplitudes according to the relation: $\mathbf{a}^{(out)}=U\mathbf{a}^{(in)}$, where $\mathbf{a}^{(in)}$ and $\mathbf{a}^{(out)}$ are the input and output vectors, respectively. Provided interferometers are lossless, their transfer matrices $U$ are unitary.

\begin{figure}[htbp]
    \centering
    \includegraphics[width=0.45\textwidth]{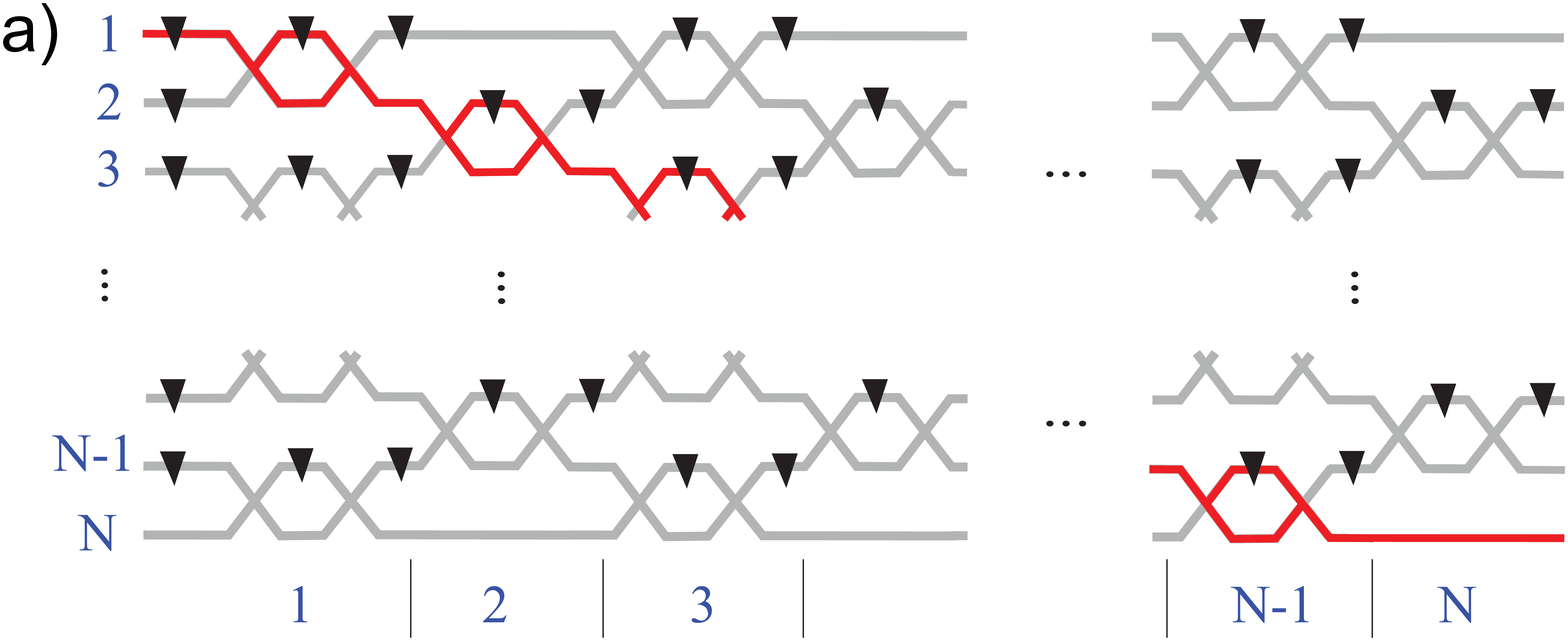}
    \includegraphics[width=0.45\textwidth]{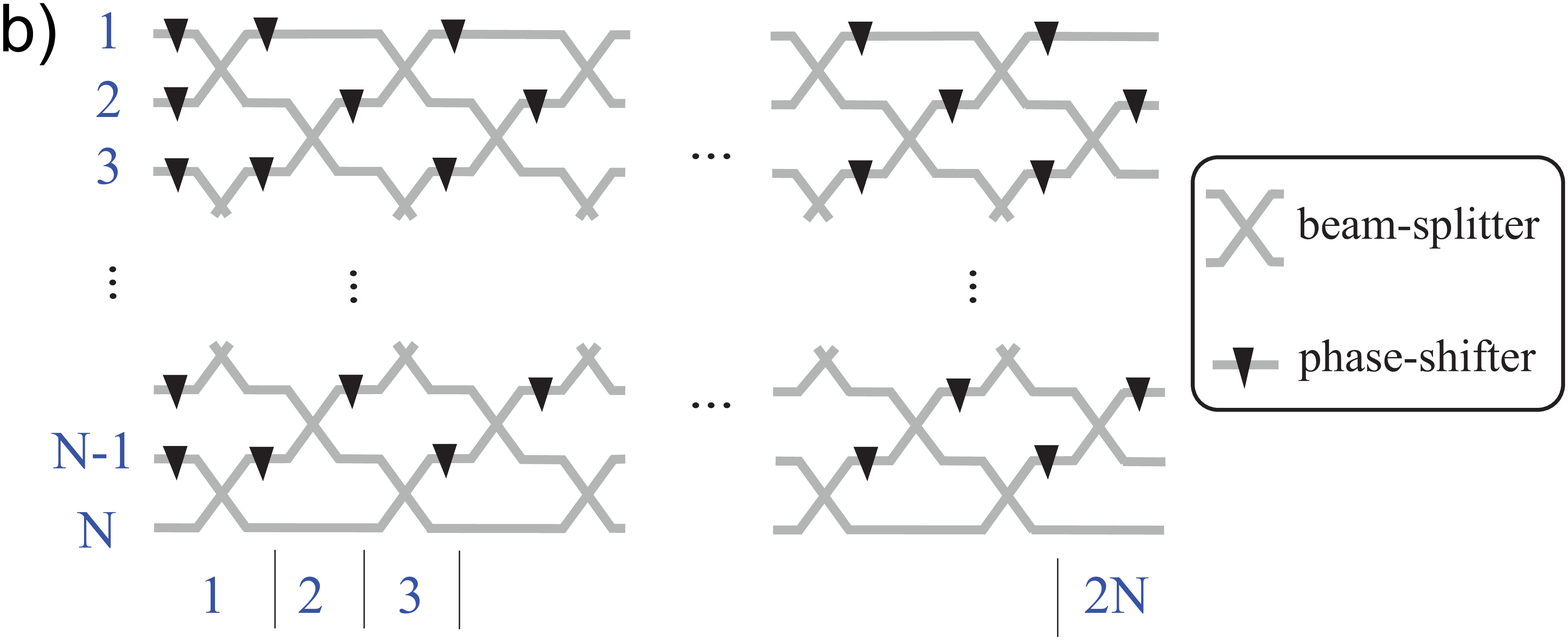}
    \caption{Two designs of  multiport interferometers:    
    a) the conventional universal MZI-based design proposed in \cite{Clements2016}. The elements colored in red are utilized in the rerouting operation from the first into the last port.
    b) the alternative BS-based design proposed in this work.}
    \label{fig:fig1}
\end{figure}

We first describe interferometers constructed with the canonical MZI-based design~\cite{Clements2016} depicted in  Fig.~\ref{fig:fig1}a. It is formed by $N$   layers consisting of MZIs, each acting  locally only on two neighboring channels. In this scheme, the overall number of MZIs is equal to $N(N-1)/2$. Accounting for the  $N-1$ phase shifts at the input, the total number of phase shifts in the interferometer is $N^2-1$, exactly the number of independent real parameters that parametrize an arbitrary unitary $N\times{}N$ matrix. 

Therefore, the transfer matrix of the MZI-based interferometer can be written as 
	\begin{equation}\label{eqn:ClementsDesign}
		U_{\text{MZI}}=V_{\text{MZI}}^{(N)}\cdot\ldots\cdot{}V_{\text{MZI}}^{(2)}V_{\text{MZI}}^{(1)}\Phi
	\end{equation}
where $V_{\text{MZI}}^{(m)}$ is the transfer matrix of $m$-th layer,
$\Phi=\text{diag}\left(e^{i\varphi_1^{(out)}},\ldots,e^{i\varphi_{N-1}^{(out)}},1\right)$ is the diagonal matrix  with 
$\varphi_j^{(out)}$ being the phase-shifts introduced in the end. In \eqref{eqn:ClementsDesign}, layer transfer matrices $V_{\text{MZI}}^{(m)}$ are of the form: $V_{\text{MZI}}^{(m)}=\prod_{j\in{}\Omega_m^{\text{MZI}}}T^{(m)}_{\text{MZI},j}(\varphi^{(m)}_{2j-1},\varphi^{(m)}_{2j})$, where 
\begin{equation}\label{eqn:MZIblock}
	T^{(m)}_{\text{MZI},j}(\varphi_{1},\varphi_{2})=\left(
		\begin{array}{ccllcccc}
		1 &  \cdots & \cdots & \cdots & \cdots &  0 \\
		\vdots   & \ddots  &  &  &  &   \vdots \\
		\vdots   &   & B_{1,1}^{(m)} & B_{1,2}^{(m)} &  &   \vdots \\
		\vdots   &   & B_{2,1}^{(m)} & B_{2,2}^{(m)} &  &   \vdots \\
		\vdots   &  &  &  & \ddots &   \vdots\\
		0  &  & \cdots & \cdots  &  &  1 
		\end{array}			
	\right)
\end{equation}
is the block matrix of single MZI placed in the $m$-th layer between channels $j$ and  $j+1$, $\Omega_m^{\text{MZI}}$ denotes the ordered sequence of MZIs in the layer with index $m$. Block matrix \eqref{eqn:MZIblock} has all diagonal elements $1$ except those labeled  $B_{1,1}^{(m)}=e^{i(\varphi_{1}^{(m)}+\varphi_{2}^{(m)})}a_j^{(m)}$ and  $B_{2,2}^{(m)}=-e^{i\varphi_{1}^{(m)}}a_j^{(m)*}$, and all off-diagonal elements equal to $0$ except  those labeled $B_{1,2}^{(m)}=e^{i\varphi_{1}^{(m)}}b_j^{(m)}$ and $B_{2,1}^{(m)}=e^{i(\varphi_{1}^{(m)}+\varphi_{2}^{(m)})}b_j^{(m)*}$, where we introduced the shorthand notations: $a=\sin\varphi_{1}^{(m)}\cos(\alpha_{1}^{(m)}-\alpha_{2}^{(m)})+i\cos\varphi_{1}^{(m)}\sin(\alpha_{1}^{(m)}+\alpha_{2}^{(m)})$  and $b=\cos\varphi_{1}^{(m)}\cos(\alpha_{1}^{(m)}+\alpha_{2}^{(m)})+i\sin\varphi_{1}^{(m)}\sin(\alpha_{1}^{(m)}-\alpha_{2}^{(m)})$. The variable phase shifts $\varphi_j$ are used to reconfigure the interferometer and have required ranges from $0$ to $2\pi$. Parameters  $\alpha_{l}^{(m)}$ describe errors caused by the imbalances of the static BSs due to non-ideal realization. 

When $\alpha_{l}^{(m)}=0$ the MZI-based interferometer is capable of implementing an arbitrary unitary transfer matrix, however,  imbalances $\alpha$'s undermine its universality.
A trivial example of rerouting from port $1$ into port $N$ of an $N$-port interferometer  is shown in Fig.~\ref{fig:fig1}a. Obviously, to attain this transformation all diagonal MZIs  should be in the cross state. However, the unit transmissivity of an MZI: $\tau=|b|^2=\cos^2\varphi_1\cos^2(\alpha_1+\alpha_2)+\sin^2\varphi_1\sin^2(\alpha_1-\alpha_2)$, can be obtained only when $\alpha_1=\alpha_2=0$.  

The schematic representation of our design is depicted in Fig.~\ref{fig:fig1}b. Our design has rectangular placement of building blocks, each of which is a single static BS and single tunable phase shifter. Hence, the name BS-based for our design. The $N$-port inteferometer of the BS-based design has $2N$ layers  so that both the scheme depth, as quantified by the maximum static BSs crossed by the signals, and the number of phase shifts are equal to those of the MZI-based design. The interferometer transfer matrix takes the form:
	\begin{equation}\label{eqn:OurDesign}
		U_{\text{BS}}=V^{(2N)}_{\text{BS}}\cdot\ldots\cdot{}V^{(2)}_{\text{BS}}V^{(1)}_{\text{BS}}\Phi,
	\end{equation}
in which layer transfer matrices $V_{\text{BS}}^{(m)}=\prod_{j\in{}\Omega_m^{\text{BS}}}T^{(m)}_{\text{BS},j}(\varphi^{(m)}_{j})$ has the block $T^{(m)}_{\text{BS}}(\varphi^{(m)})$ of form  \eqref{eqn:MZIblock}, but with $B_{1,1}^{(m)}=e^{i\varphi^{(m)}}\cos(\theta_0+\alpha^{(m)})$, $B_{2,2}^{(m)}=\cos(\theta_0+\alpha^{(m)})$, $B_{1,2}^{(m)}=\sin(\theta_0+\alpha^{(m)})$ and $B_{2,1}^{(m)}=-e^{i\varphi^{(m)}}\sin(\theta_0+\alpha^{(m)})$. Here, angle $\theta_0$ quantifies the transmission of the static BSs, which specific value will be given below.

\begin{figure}[htbp]
    \centering
    \includegraphics[width=0.22\textwidth]{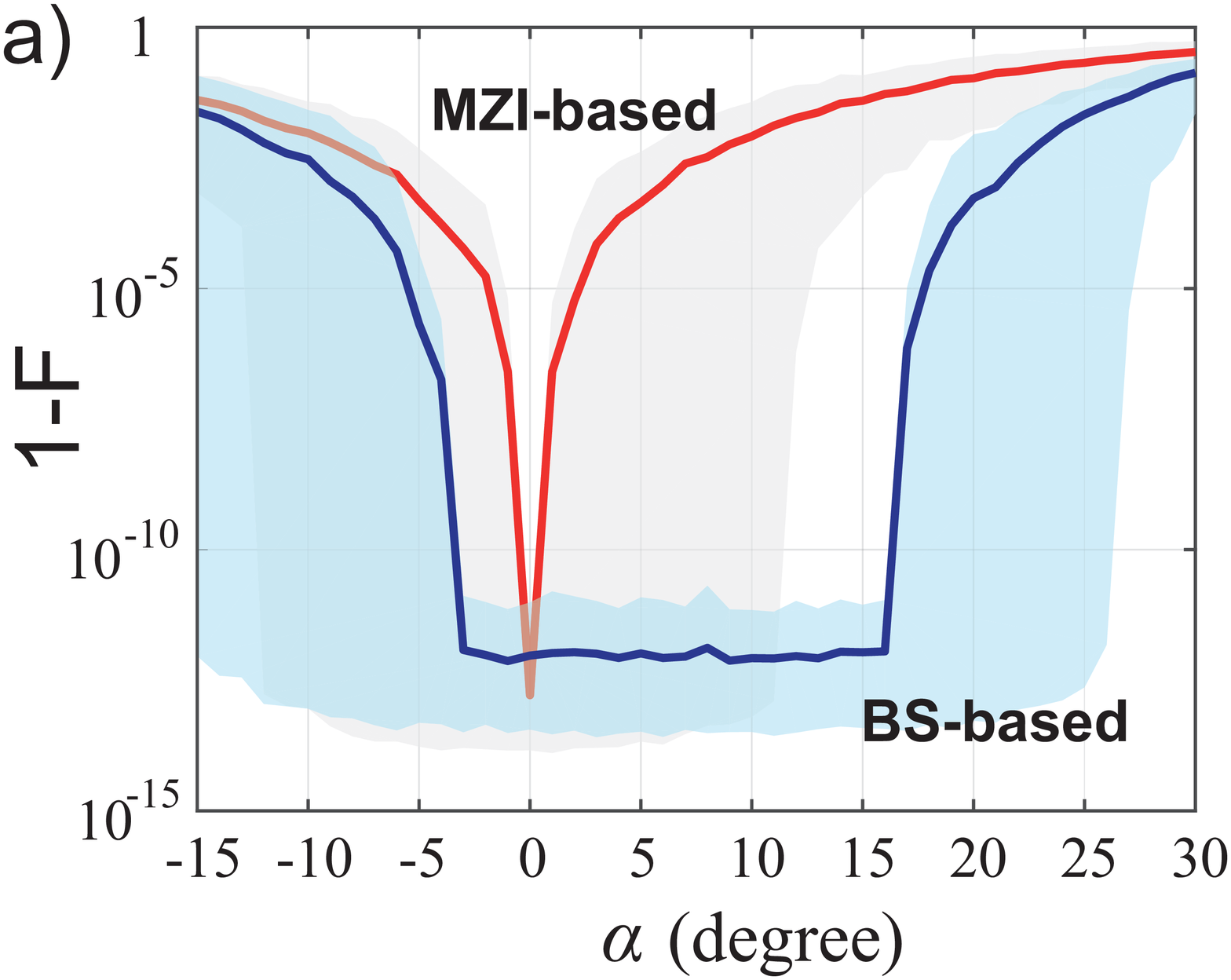}
    \includegraphics[width=0.24\textwidth]{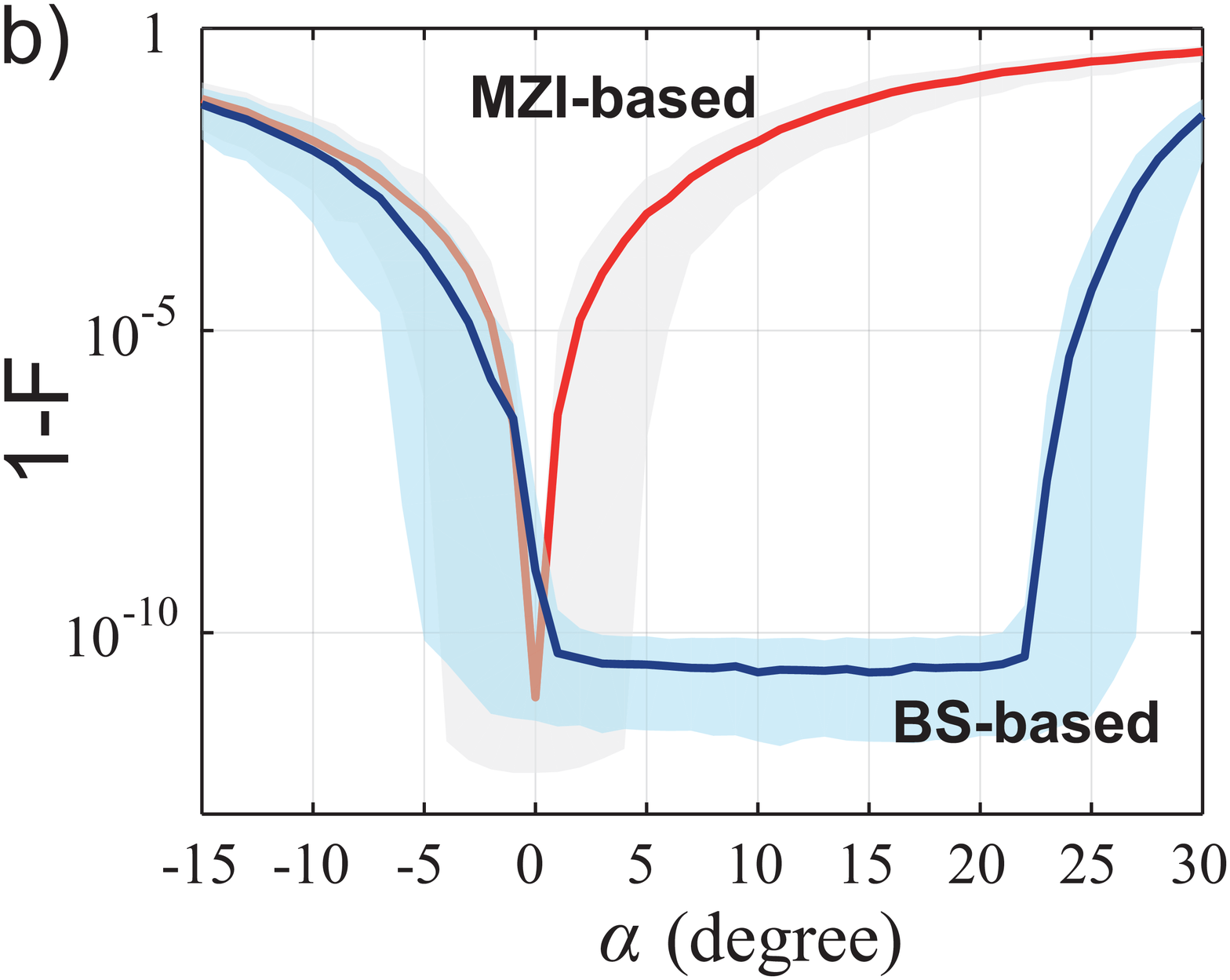}
    \caption{Infidelity $1-F$  as a function of error parameter $\alpha$ for the MZI-based and BS-based interferometers at $\theta_0=\pi/4$ and a) $N=5$ and b) $N=10$. For each value of $\alpha$ the infidelity distribution were obtained numerically using a set of $300$  unitary matrices drawn randomly from uniform distribution. The solid curves correspond to the average over all samples size; the lower and upper boundaries of the shaded regions are averages for $10$ infidelities with the lowest and highest values, respectively. }
    \label{fig:fig2}
\end{figure}
\begin{figure}[htbp]
    \centering
    \includegraphics[width=0.22\textwidth]{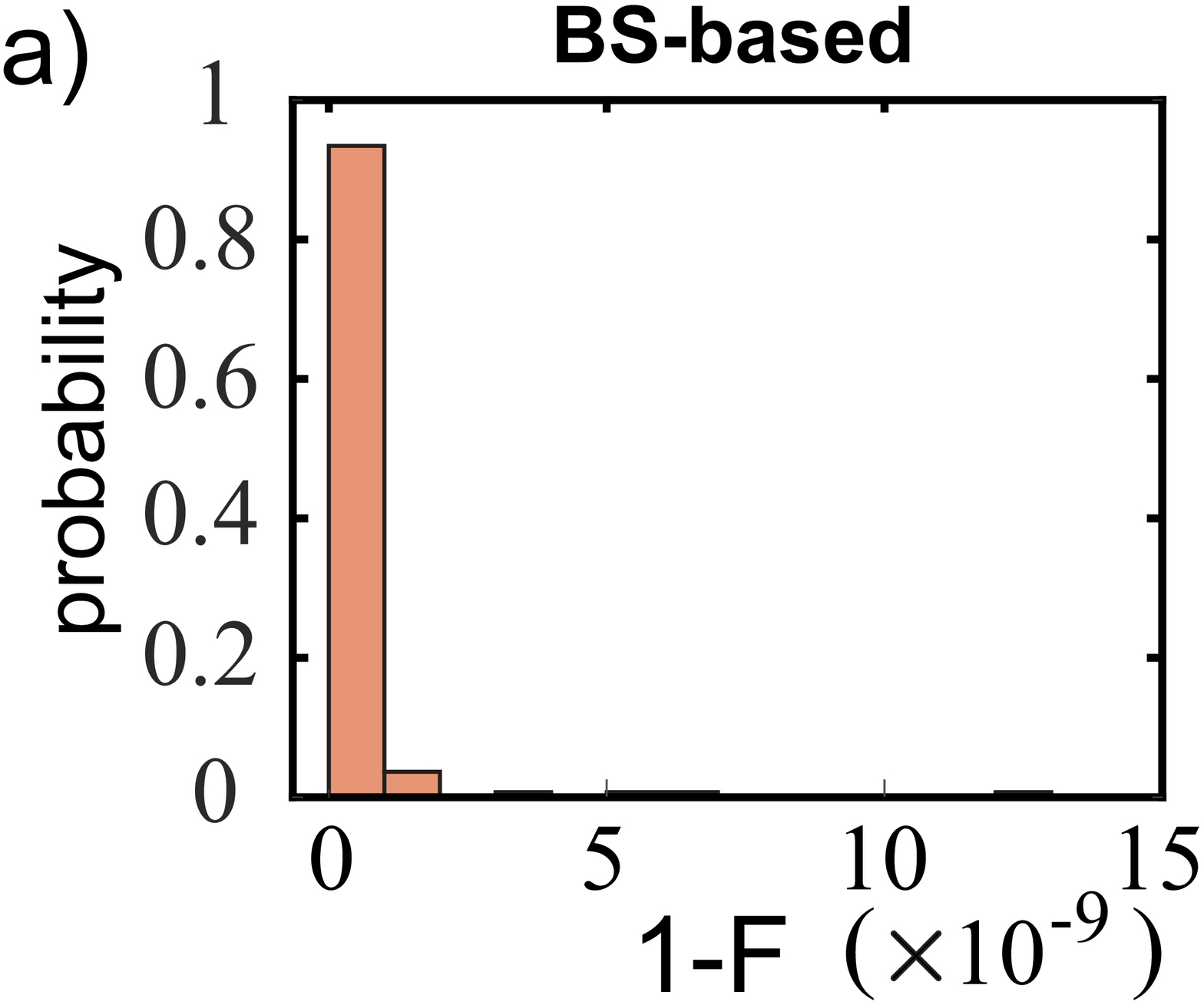}\hspace{10pt}
    \includegraphics[width=0.22\textwidth]{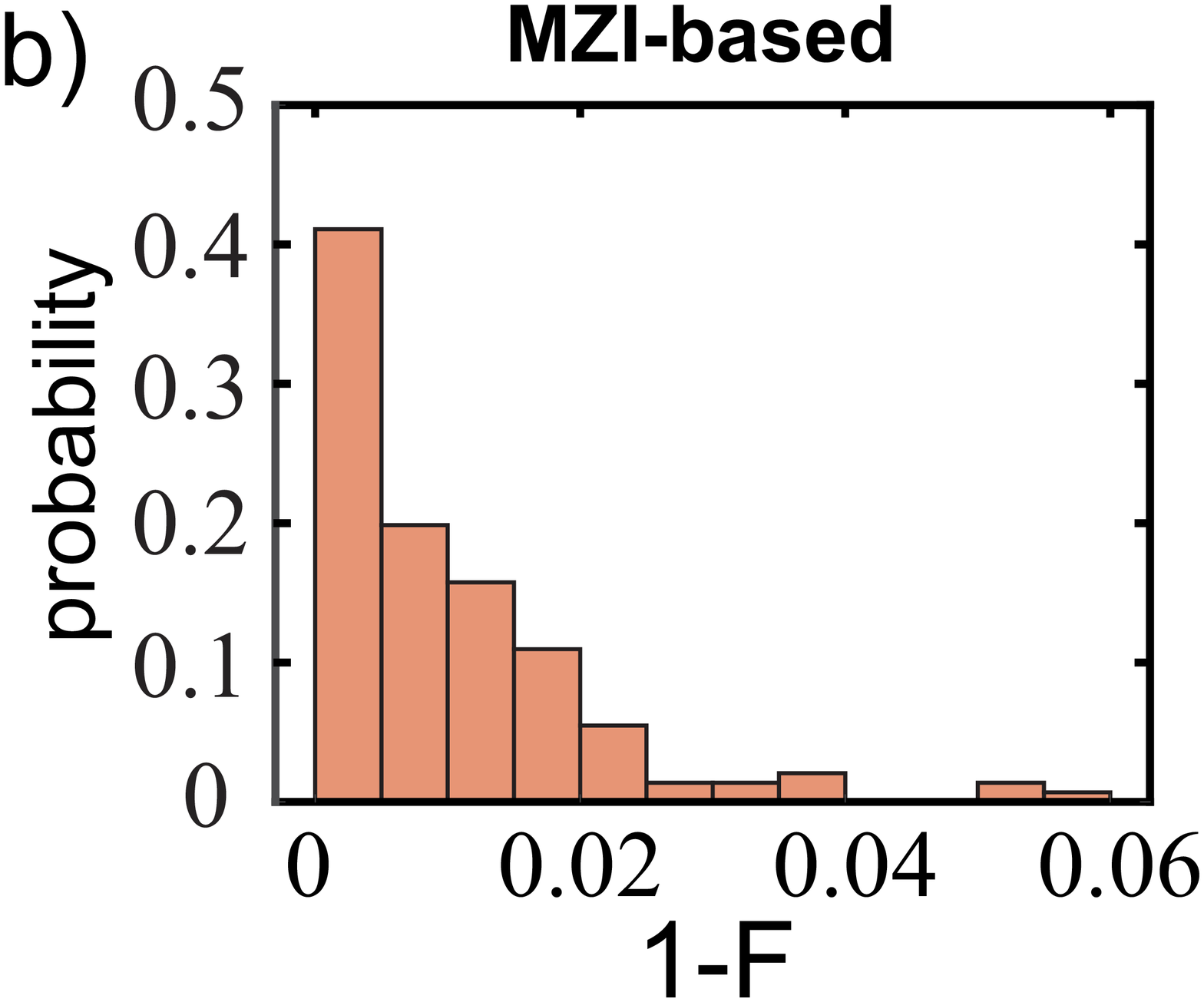}
    \caption{Normalized histogram of infidelity $1-F$ for the BS-based (a) and MZI-based (b) $10$-port interferometers at random errors. The error angles $\alpha_j$ were drawn from the  distribution $p(\alpha)=\exp\left(-\alpha^2/2\Delta^2\right)/\sqrt{2\pi}\Delta$ with $\Delta=10$ degrees. For the BS-based interferometer the parameter  $\theta_0=55$ degrees, roughly corresponding to the center of the high-fidelity plateau, depicted in Fig.~\ref{fig:fig2}b.  The histograms is the result of the optimization of $300$ randomly sampled unitary matrices.}
    \label{fig:fig3}
\end{figure}

\section{Error tolerance}

We consider errors as $\alpha_j\ne{}0$ that tune the splitting ratios of the static BSs off required values. Firstly, we study the effect of coherent errors at which $\alpha_j=\alpha$. For the schemes manufactured by planar lithography techniques this type of errors is linked to the variations of waveguide's material and geometry, which is dominated as their spatial scale is usually large compared with the area occupied by the scheme~\cite{BogaertsChrostowski}. These arguments  can also be applied to interferometers manufactured by other methods, for example, femtosecond direct laser writing \cite{Dyakonov2018}, as well as alternative implementation approaches exploiting repetitively few optical elements to obtain the desired transformation between multiple modes~\cite{RohdeLoop}.

We evaluate the performance of multiport interferometers by calculating the fidelity, defined as:
\begin{equation}\label{eqn::fidelity}
    F=\frac{\left|\mathrm{Tr}(U^{\dagger}U_{0})\right|^2}{N\mathrm{Tr}(U^{\dagger}U)},
\end{equation}
which compares the target unitary matrix $U_0$ and the actual transfer matrix $U$ realized by the interferometer, where $N$ is the size of the matrices. Provided that the matrices $U$ and $U_0$ are equal up to a complex multiplier, the fidelity \eqref{eqn::fidelity} gets its maximum value of $F=1$.

Generally, no analytical solution is known to derive phase shifts that maximize the fidelity \eqref{eqn::fidelity}, except for the case of error-free MZI-based interferometers, for which an analytical procedure is provided in \cite{Clements2016}. Unfortunately, we could not found analogous procedure for the error-free BS-based interferometers. 

We used a numerical optimization algorithm based on the basinhopping algorithm. Given a unitary matrix $U_0$, the algorithm was searching for a global minimum of infidelity $1-F$ over the space of phase-shifts. To decrease the chance of falling into local minima, we used multiple runs of the optimization with random initial values of the phases. Each numerical experiment involved  optimization over a series of target matrices $U_0$,  drawn from the Haar random distribution using the method based on the QR-decomposition of random matrices~\cite{Mezzadri2006}. With this algorithm it took several hours to find optimal phase shifts for a single $10\times{}10$ transfer matrix, so that a multi-core computer has been utilized to derive required dependencies. We understand that more efficient numerical algorithms can be  developed for this specific task~\cite{MillerOptimization}.

The obtained infidelity $1-F$ as a function of the error parameter $\alpha$ is plotted in Fig.~\ref{fig:fig2}. The finite accuracy of the numerical algorithm sets the minimal infidelity value of $\sim{}10^{-12}-10^{-10}$, which could not be overcome neither for the MZI-based interferometer with $\alpha=0$ where exact zero was expected. The MZI-based interferometers are equally sensitive to both positive and negative values of  $\alpha$ with the acceptable range of errors is of the order of several degrees.
For the BS-based interferometers, infidelity behaves radically different: while at $\alpha<0$  the performance of the two are comparable, at $\alpha>0$ the BS-based interferometers provide perfect fidelity for $\alpha$ as large as $\sim{}20$ degrees --- several times larger than for the MZI-based interferometers. 
Fig.~\ref{fig:fig2} suggests that  when the positive and negative values of  $\alpha$ are equiprobable, the optimal choice of the static BSs defined at the design stage is such that $\theta_0\approx{}55$ degree, corresponding to the center of the high-fidelity plateau.

The superior perfomance of the BS-based interferometers at positive $\alpha$'s can be attributed to the interplay of two competing properties. On the one hand, more transmissive BSs with $\alpha>0$ enable more efficient travelling of the signal amplitudes across the scheme, which  has been recognized as a prerequisite for robustness to errors in other universal interferometer architectures~\cite{RobustArchitecture2019}.  On the other hand, the increase of the transmissivity reduces the beam-splitter interaction, which is completely absent in the limiting case of $\alpha=\pi/4$.

Secondly, we consider the incoherent errors at which $\alpha$'s are distributed at random across the scheme. The obtained infidelities are shown in Fig.~\ref{fig:fig3}. Clearly, the BS-based interferometer design is ultimately tolerant to the incoherent errors.

\begin{figure}[htbp]
    \centering
    \includegraphics[width=0.45\textwidth]{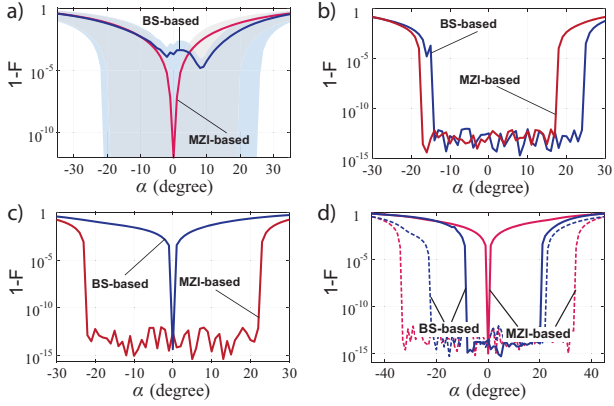}
    \caption{ Infidelity $1-F$ as a function of error parameter $\alpha$ for the BS- and MZI-based interferometers implementing a) $300$ randomly generated unitary $U_0$ at $N=3$; the lower and upper boundaries of the shaded regions are averages for $10$ infidelities with the lowest and highest values, respectively.  b)  DFT matrix at $N=3$, c) the swap between port $1$ and $2$ at $N=3$, d)  Hadamard transformation $H_2$ of ports $1$ and $2$ at $N=3$ (dashed curves) and  Hadamard transformation $H_4$ at $N=4$ (solid curves). }
    \label{fig:fig4}
\end{figure}

The performed analysis cannot be complete in proving strict universality of the interferometers, since random generation of target matrices can overlook small subsets that are non-implementable error-tolerantly or non-implementable at all. We now show that such  matrices do exist for the BS-based design. For this, we generate random matrices of size $N=3$ and calculate infidelities. Fig.~\ref{fig:fig4}a demonstrates that in this low-dimensional case the relative volume occupied by the non-implementable matrices is quite large to be caught by random sampling:  $\approx{}6$ per cent of them satisfy the condition $1-F>10^{-5}$ at $\theta_0=\pi/4$, $\alpha=0$.
However, taking the results of Fig.~\ref{fig:fig2} into account, where not a single non-implementable matrix was present, we conclude that the relative volume of non-implementable matrices rapidly diminishes with $N$.

Also, we consider some concrete examples which fall out of the picture obtained with random sampling. One example is of the discrete Fourier transform (DFT) matrix with elements $U^{(DFT)}_{mn}=\exp(i2\pi(m-1)(n-1)/N)/\sqrt{N}$. Fig.~\ref{fig:fig4}b illustrates the resutls for $N=3$, showing that both designs are equally tolerant, which is actually true for larger $N$. 

We next consider a block-diagonal matrix of size $N=3$ that comprises the swap block of ports $1$ and $2$ (Fig.~\ref{fig:fig4}c). In contrast to previous results, the MZI-based design works much better than the BS-based design. Moreover, we have found that the block-diagonal matrices of arbitrary sizes are not reproduced error-tolerantly  by the BS-based schemes, thus, it represents a class of transformations where the MZI-based design is advantageous.

Finally, we consider the Hadamard transformation. Namely, matrix $H_2=\frac{1}{\sqrt{2}}\left(\begin{array}{cc}1&1\\-1&1 \end{array}\right)$  as a constituent upper-left block of a $3$-by-$3$ matrix and matrix $H_4=H_2\otimes{}H_2$  as a whole $4$-by-$4$ matrix. As can be seen from Fig.~\ref{fig:fig4}d, both designs are good at implementing the transformation when it is a part of a larger matrix, while the BS-based design is much better in case of a whole martix. Once again, this is the evidence that the block-diagonal matrices are better realized by the MZI-based  than BS-based interferometers. In addition, we have found the permutation matrices are also better implemented by the MZI-based interferometers rather than the BS-based ones. In particular, rerouting operation depicted in Fig.~\ref{fig:fig1}a cannot be implemented perfectly by the BS-based interferometer.

\section{Loss tolerance}

\begin{figure}[htbp]
    \centering
    \includegraphics[width=0.2\textwidth]{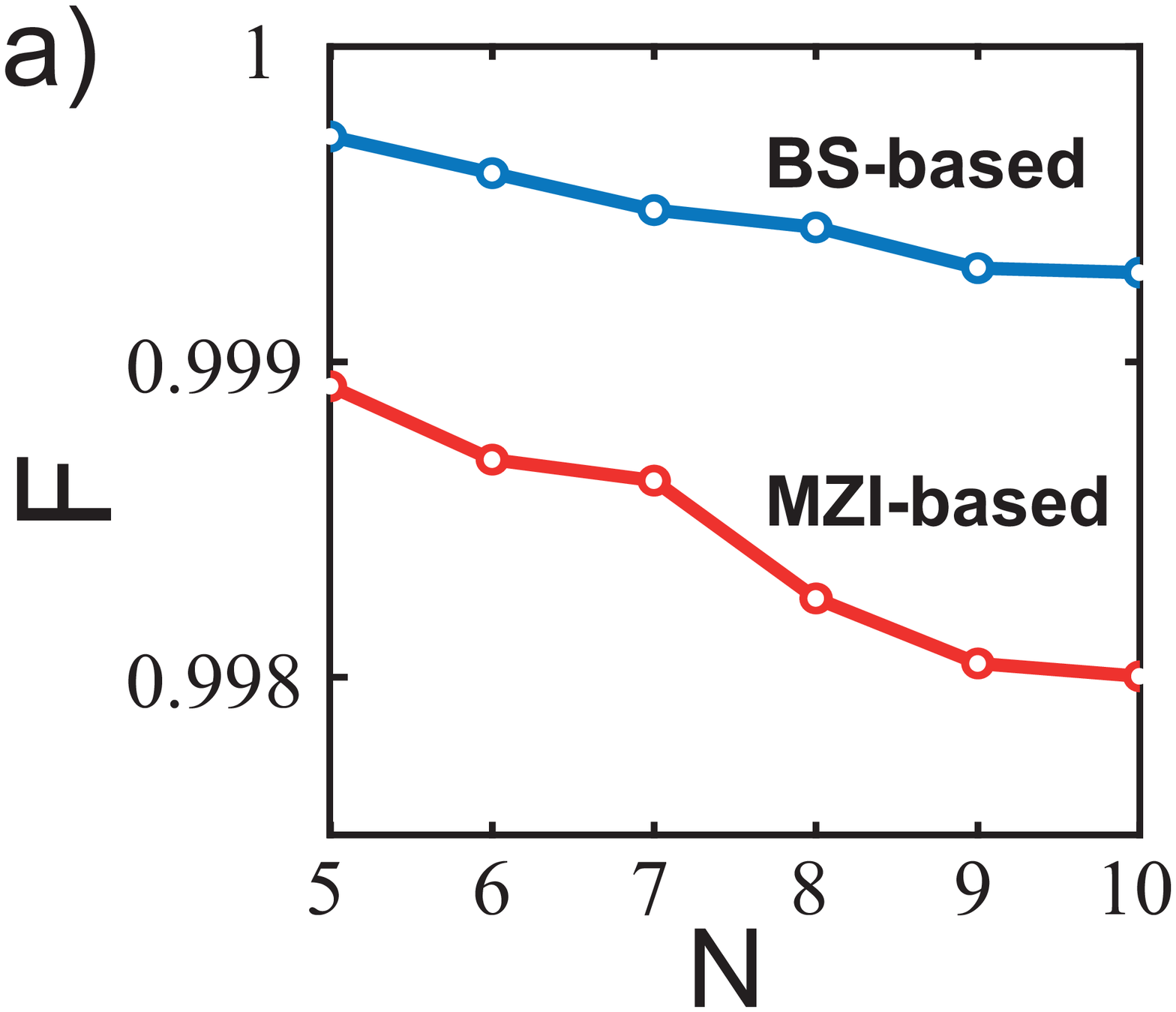}
    \includegraphics[width=0.25\textwidth]{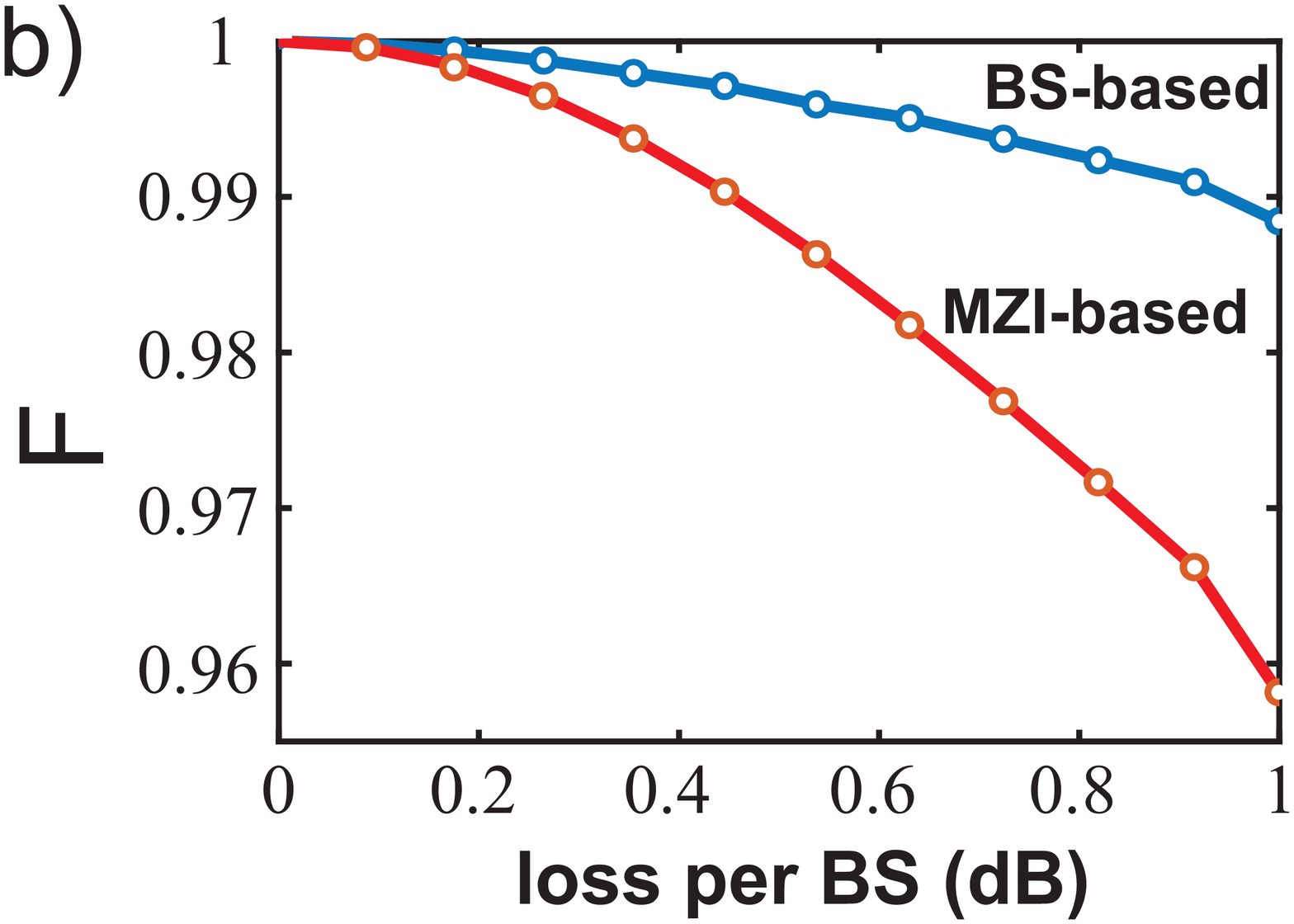}
    \caption{a) average fidelity at a loss of $0.2$ dB per static BS for different interferometer sizes $N$, b) average fidelity of a $10$-port interferometer as a function of loss introduced by each static BS. In the BS-based design $\theta_0=55$ degrees, while BSs in the MZI-based design  are balanced.  Each average was calculated over $50$ randomly sampled matrices.  }
    \label{fig:fig5}
\end{figure}

In addition, we study the effect of unbalanced losses that occur because different paths through the interferometer experience different losses, which can result in poor fidelity. Typically, static  BSs introduce additional losses due to waveguide bending and scattering, therefore, in both designs, unbalanced losses occur when the signals have to pass through side paths, as they contain less BSs than the inner paths. To model lossy BSs, each $B_{i,j}^{(m)}$ in the MZI-based and BS-based transfer matrices was multiplied by $t^2$ and $t$, respectively, where $0\le{}t\le{}1$ is the BS transmission coefficient. Then, random target matrices were generated and the global maxima of fidelity \eqref{eqn::fidelity} were calculated. The obtained results are shown in Fig.~\ref{fig:fig5} suggesting that the BS-based design is more loss-balanced than the MZI-based design. This can be attributed to more evenly distributed static BSs in the scheme than in the MZI-based design, where BSs are grouped in pairs. We understand that transformations might exist where our design loose the advantage, however, we have not found any outside of the set of non-implementable examples described above.

\section{Conclusion}

In summary, we proposed a novel design of programmable multiport interferometers, which exhibit superior tolerance to errors and losses than the previously known counterparts, while not requiring redundant elements. It is noteworthy that the proposed design is less complex than the counterparts with possible implementation by a variety of experimental platforms. For example, the static interferometers that have been used  in experiments on boson sampling \cite{PanBosonSampling, CrespiAndersonLocalization} have suitable placement of the passive BS elements, yet lacking programmability.

\medskip

\noindent\textbf{Funding.} The reported study was funded by RFBR according to the research project No 19-52-80034.




\bibliography{bibl}


\end{document}